\documentstyle[11pt,epsf]{article}
        \def\be{\begin{equation}}
        \def\ee{\end{equation}}
\def\bea{\begin{eqnarray}}
\def\eea{\end{eqnarray}}
\def\l{\label}
\def\r{\ref}
\def\c{\cite}
\begin{document}
\begin{center}
{\LARGE{\bf Inhomogenous Poisson Networks
               and Random Cellular Structures}}\\
\vspace{3cm}
 {\bf $^{\dag, \ddag}$Kh. Saaidi \footnote{ E-mail: ksaaidi@hotmail.com},
 $^{\dag, \ddag}$ M.R. Setare  \footnote{E-mail: rezakord@yahoo.com}
 }\\
 {\it${^{\dag}}$ Department of Science, Kurdistan  University,
 Pasdaran Ave., Sanandaj, Iran} \\
 {\it $ ^{\ddag}$ Institute for Studies in Theoretical Physics and Mathemaics,
 P.O.Box, 19395-5531, Tehran, Iran}\\

\vspace{2cm}
{\bf Abstract}\\
\end{center}

we study the statistical properties of inhomogenous Poisson
networks. we perform a detailed analysis of the statistical
properties of Poisson networks and show that the topological
properties of random cellular structures, can be derived from
these models of random networks. we study both two and three
dimensional networks with non uniform density and show that with a
class of symmetric distribution $P(\lambda_{1},
 \lambda_{2}, \ldots , \lambda_{N})$
Lewis and Aboav-Weaire laws are obeyed in these networks.\\
\newpage
{\section {Introduction}}

In the past three decades there has been a considerable interest
in studying a class of non-equilibrium systems known collectively
as {\it Cellular Structures}\c{S1}. The geometrical and dynamical
properties of these systems are best displayed in the familiar
pattern formed by a soap froth confined between two transparent
plates. Other examples include  polycrystalline domains in metals,
ceramics, magnetic domains, biological tissues and mono layers of
fatty acids on surface of water. One can also mention other
examples from material science, like cracking in glazes, fracture
and dewetting  of polymer films above their glass transition
temperature \c{S2}-\c{Le1}. Despite the diversity in systems in
which cellular structures  are formed, numerous experiments have
shown that the long time statistical behavior of these systems are
characterized by  certain universal, system independent laws. This
means that topological  and geometrical constrains influence the
properties of these networks in a very essential way. For example,
the fact that energy is associated with the length of the edges of
cells, immediately leads to the conclusion that in the two
dimensional  structures all the vertices are 3-valent. Combination
of this result with the Euler character formula shows that the
average number of sides per cell, $\langle i \rangle$ is equal to
6: \be\l{1} \langle i \rangle=6. \ee Similar considerations in
3-dimensions, where energy is associated with the area of faces of
cells, proves that all vertices are 4-valent and that: \be\l{2}
   \langle f \rangle = \frac{12}{6-\langle i \rangle},
\ee
where $\langle f \rangle$ and $\langle i \rangle$ are  the  average
number of faces per cell and the average number of edges per face
 respectively. Besides these properties which are a direct consequence of
 the Euler character formula, experiments have revealed a number of very
 general, properties among which the most important are : i) Von-Neumann \c{vo} ,
ii) Lewis \c{La} and iii) Aboav-Weaire law \c{Ab}. Von-Neumann law refers to
the rate of expansion of a single cell and is accounted for theoretically
in a satisfactory way. It has also been generalized to curved surfaces where
the curvature of the surfaces plays a role both in the dynamic of a single
cell and in it's stability \c{Av}. The Aboav-Weaire law which is statistical in nature
and refers to the correlation of adjacent cells has been shown to hold in
random 3-valent graphs, viewed as planar Feynman graphs of $\phi^3$ theory
and solved by techniques of matrix models \c{Go}. The authors of \c{As} have
investigated the statistical properties of two dimensional random cellular
systems in term of their shell structures.\\
However there has been no explanation of the empirical Lewis law which states
that the average area of cells has a linear relation with the number of cells,
for large number of edges. However recently there have
been some attempts \c{F1} to derive both Lewis
and Aboav-weaire laws, from Poisson networks, i.e. networks based on a Poisson
distribution of horizontal segments between a fixed set of parallel lines.
In ref.\c{F1} it was shown that in such networks both the Aboav-Weaire and the
Lewis laws were obeyed. The analysis of ref. \c{F1} was however restricted to
the uniform density case, i.e: where the distance between the horizontal lines
were all equal. In this paper we extend this analysis to the non-uniform
case and add one more ingredient of randomness to the network, where the
distribution of vertical lines is also probabilistic.
The paper consists of two parts, the first part (section 2), where we
rederive the results of \c{F1} in a somewhat clearer way, and part two (section
3) where we extend these results to the non-uniform density case.

{\section{ Poisson Networks With Uniform Density}}

The two dimensional Poisson networks are generated as follows (fig.1)\c{F1}.
One takes a family of parallel lines in the y-direction in the plane.
The distance  between lines does not affect the  topological
properties of the network, although it affects the geometry. This
distance is taken as uniform and equal to $d_0$. Suppose there are N
columns, $C_{1}, C_2, \ldots, C_{N}$, in the network. The region between
two successive columns is divided into cells.
The division of each column $C_{\alpha}$, is based on a Poisson point
distribution in the y-direction. An edge in the x-direction is
taken through each Poisson point (P-point). In this way a 3-valent
network is generated in the plane which, although different from the
realistic cellular structures, is simple enough to be analyzed closely for
the study of various statistical-topological properties of the network.\\
\begin{figure}[t]
\begin{center}
\leavevmode
\epsfxsize=80mm
\epsfysize=40mm
\epsfbox{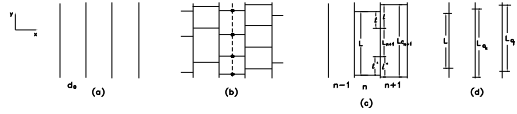}
\caption{{\it 2D Poisson network. (a) Parallel columns in the $x$ direction
with width $d_0$. (b)These columns are divided into cells by edges through
the points of independent uniform or non-uniform poisson distributions.
The structure in c is used for obtaining  the probability  density
of cover lengths, $L_{c}$, for a given $L$.(d)Cover lengths of segment $L$
in two distribution which partly overlap.}}
\end{center}
\end{figure}

The 3-dimensional tetravalent network are generated by a similar
process (fig.2). One takes an arbitrary 2-dimensional trivalent
network(base network) on the xy plane (fig.2.a) and takes vertical planes
(parallel to the z-axis) through each edge  which  divides the 3-dimensional
space into prismatic columns. In each column one considers a Poisson
distribution of uniform density in the z-direction  and divides the
columns into cells by planes perpendicular to the z-axis through each P-point
(fig.2.b). The height of a cell, $L$, is the distance between
adjacent P-points in the associated distribution (fig.2.c). The average
cell height is unity.\\
\begin{figure}[t]
\begin{center}
\leavevmode
\epsfxsize=80mm
\epsfysize=40mm
\epsfbox{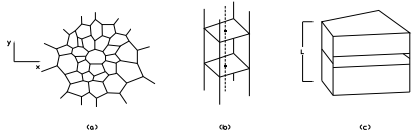}
\caption{{\it Three dimensional Poisson network. (a) an arbitrary
planar network. (b)The columnar structure of three dimensional Poisson network.
The cell in $c$ has $i=4$, $F=10$, and height $L$. }}
\end{center}
\end{figure}
Consider one of the columns, say $C_{\alpha}$. The Poisson distribution of
$n$ points in a segment of length $L$ is given by :
\be\l{3}
  P(n)=\frac{L^n}{n!} e^{-L}.
\ee
Consider a particular set of columns $C_{\alpha}, C_{\beta}, \ldots ,C_{\gamma}$.
The probability that in a given distance $L$, there are $n_{\alpha},n_{\beta},
\ldots ,n_{\gamma}$ segments respectively in columns $C_{\alpha}, C_{\beta},
\ldots , C_\gamma$, is equal to
\be\l{4}
\Phi(L, n_\alpha, n_\beta, \ldots , n_\gamma) \equiv
P(n_\alpha) P(n_\beta) \ldots  P(n_\gamma).
\ee
Now consider a reference column which we denote by $C_{\alpha}$. The probability
that there is a cell with height between $L$ and $L+dL$ in this column is
       $e^{-L}dL$.
I denote the columns which are neighbors of this reference column by
$C_{\alpha_1}, C_{\alpha_2},\ldots , C_{\alpha_k}$. The number of these neighboring
columns depend on the dimensionality and type of the network. The joint probability
that there is a cell of height between $L$ and $L+dL$ in the column $C_\alpha$
 such that its neighbors $C_{\alpha_1}, C_{\alpha_2},\ldots , C_{\alpha_k}$
  respectively have $ n_{\alpha_1}, n_{\alpha_2}, \ldots , n_{\alpha_k}$ points in the interval $L$ is :
\be\l{5}
\Psi_{\alpha}(L, n_{\alpha_1}, n_{\alpha_2}, \ldots , n_{\alpha_k})dL =
P(n_{\alpha_1}) P(n_{\alpha_2}) \ldots  P(n_{\alpha_k}) e^{-L}dL.
\ee
Clearly this distribution function depends on the individual values of the
variables  $n_{\alpha_1}, n_{\alpha_2}, \ldots , n_{\alpha_k}$, however
in the sequel another probability distribution will be useful. That is
$\Psi(L,I)dL$ were I is defined as $I= n_{\alpha_1}+n_{\alpha_2}+\ldots +
 n_{\alpha_k}$. The joint probability $\Psi(L,I)dL$, is calculated as follows:
 \be\l{6}
 \Psi_\alpha(L,I)dL= \sum' \Psi_{\alpha}(L,n_{\alpha_1},n_{\alpha_2},
 \ldots,n_{\alpha_k})dL=\frac{(kL)^I}{I!}e^{-(k+1)L}dL,
 \ee
where $ \sum' $ means the sum over all $ n_{\alpha_i } $'s subject to
the constraint that their sum be equal to I.
It's important to note that the distribution $\Psi_\alpha(L,I)$ is obtained
from a sum of distribution of the form $\Psi_{\alpha}(L,n_{\alpha_1},n_{\alpha_2},
\ldots,n_{\alpha_k})$ and is not equal to any of them.(compare with eq.(9a)
of\c{F1}).\\
\subsection { Statistical Properties of Poisson Networks}
{\Large 2.1.1 Two Dimensional Networks}

In two dimensions, every column has two neighbors, hence in eq.(\r{6}), $k=2$, and
the number of edges of a cell (see fig.1) in a reference column is equal to :
\be\l{7}
i=I+4.
\ee
Combining eqs.(\r{6}) and (\r{7}) gives the probability distribution $g(L,i)$ of
finding {\it i-cells} of height in the interval $(L, L+dL)$:
\be\l{8}
g(L,i)dL=\frac{(2L)^{(i-4)}}{(i-4)!}e^{-3L}dL.
\ee
Clearly this distribution function is normalized.
\be\l{9}
\int_{0}^{\infty} \sum_{i=4}^{\infty} g(L,i)dL=1.
\ee
From eq.(\r{8}) one can obtain the average height of {\it i-cells}:
\be\l{10}
  <L>_i=\int_{0}^{\infty} L {g(L,i)\over g(i)} dL,
\ee
where $ g(i) $ is the total probability distribution of $\it{i-cells}$
which  given by:
\be\l{11}
g(i)=\int_{0}^{\infty} g(L,i)dL=\frac{1}{3}(\frac{2}{3})^{i-4}.
\ee
A simple computation now gives:
\be\l{12}
<L>_i = {i-3 \over 3}.
\ee
  This equation, then can be used to calculate the average area of
  {\it i-cells}. For the Poisson networks (fig.1),
  where the width of all cells are equal to $d_0$, we find:
  \be\l{13}
  <A>_i=d_0<L>_i=\frac{d_0(i-3)}{3},
  \ee
which as far as linear dependence on the number of edges is concerned, agrees
with Lewis law.
One can also define a generating function:
  \be\l{14}
  G(q)=<e^{iq}>\equiv\sum_{i=4}^{\infty} g(i)e^{iq},
  \ee
a simple calculation shows that:
  \be\l{15}
  G(q)=\frac{e^{4q}}{3-2e^q}, \hspace{1cm} or \hspace{1cm}
    lnG(q)=4q-ln(3-2e^q).
  \ee
  From which one obtains by successive differentiation various connected
  moments of the distribution:
  \be\l{16}
  <i>=6,
  \ee
  \be\l{17}
  <i^2>-<i>^2 = 6,
  \ee
  \be\l{18}
  <i^3>-3<i^2><i>+2<i>^3=30.
  \ee
In the remaining part of this subsection we review briefly the basic steps
of the analysis of ref.\c{F1} for derivation of the Aboav - Weaire law.
Consider an {\it i-cell} $a$ of length $L$ in a two dimensional Poisson
network(fig.3). This cell is in column $0$ and has two neighbors in this
column, called $a'$ and $a''$.\\
There are two adjacent neighbors $1$ and $2$, which respectively
distribute $n_1$ and $n_2$ points inside the cell $a$. The number of
sides of, $a$, is then equal to
\be\l{19}
i=n_1+n_2+4.
\ee
The total number of sides of the cells in column $1$ adjacent to $a$ is:
\be\l{20}
J_1=4(n_1+1)+(V_1+1)+(V'_1+1)+V''_1,
\ee
where the meaning of the numbers $ V_1$, $V_1' $ and $ V_1'' $ are specified
in fig.3. For obtaning the Aboav-Weaire law we use  one of the  three figures
 (fig.3$a$), (fig.3$b$) or (fig.3$c$) and results are equivalent.
 Clearly $im_i$ which is the total number of sides of the
cells adjacent to $a$ is:
\be\l{21}
im_i=12+J_1+J_2.
\ee
Where the definition of $J_2$ is similar to that of $J_1$ and the number
$12$ comes from the average of sides of $a'$ and $a''$. For the average
value of $ V_1$, $V'_1  $ and  $ V''_1 $ we use:
\be\l{22}
<V_1+V'_1>=\frac{<L_{c_1}-L_a>}{1}=<L_{c_1}-L_a>,
\ee
\be\l{23}
<V''_1>=\frac{<L_{c_1}>}{1}=<L_{c_1}-L_a>+<L_a>.
\ee
Where the segment $L_{c_1}$ in column $C_1$, which is the smallest segment
containing $L_a$, is called the covering  length of $L_a$ \c{F1}.
With the probability distribution found in \c{F1} for $L_c$
it is shown that the average $<L_c-L_a>=2$, in any arbitrary column,
it then follows that:
\be\l{24}
<im_i>=24+4(n_1+n_2)+<V_1+V'_1>+<V_2+V'_2>+<V''_1>+<V''_2>.
\ee
By substituting eqs.(\r{10}), (\r{22}), (\r{23}) and (\r{24}) one finds that:
\bea\l{25}
<im_{i}>&=&16+4i+2<L>_i ,\nonumber \\
&=&16+4i+\frac{2(i-3)}{3}= 14 + {14\over 3}i.
\eea
Combining eq.(\r{25}) with (\r{12}) one obtains
\be\l{26}
<im_{i}>=16+4i+\frac{2(i-3)}{3}= 14 + {14\over 3}i \ee

This is in accord with Aboav-Weaire law, i.e: $ <im_{i}>=c_1 + c_2 i $
with $ c_1$ and $c_2 $ which are 14 and $\frac{14}{3}$ respectively.\\
\begin{figure}[t]
\begin{center}
\leavevmode
\epsfxsize=100mm
\epsfysize=50mm
\epsfbox{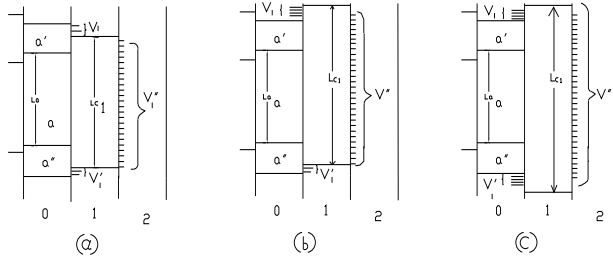}
\caption{{\it These figures are used for obtaining Aboav - Weaire
law in two and three dimensions.}}
\end{center}
\end{figure}

{\Large 2.1.2.  Three-Dimensional Networks.}

Consider a 3-dimensional network and a column whose base is an {\it i-cell.}
In eq.(\r{6}), we should now equate k with i and, relate I with the number of faces as follows:
\be\l{27}
I+2+i=F.
\ee
The probability distribution of {\it F-cells} (cells with F-faces) of height
between $L$ and $L+dL$, is now obtained as:
\be\l{28}
g_i(L,F)dL=\frac{(iL)^{F-i-2}}{(F-i-2)!}e^{-(i+1)L}dL.
\ee
The total distribution of ${\it F-cells}$ whose base is an ${\it i-cell}$
is given by:
\be\l{29}
 g_i(F) = \int_{0}^{\infty} g_i(L,F)dL=\frac{1}{(i+1)}(\frac{i}
{i+1})^{F-i-2}.
\ee
One can then obtain the average height of {\it F-cells} whose bases are
{\it i-cell}.
\be\l{30}
<L>_{i,F}=\int_{0}^{\infty}\frac{Lg_i(L,F)dL}{g_i(F)}=\frac{F-i-1}{(i+1)}.
\ee
The average height of {\it F-cells} with any base is given by:
\be\l{31}
<L>_F=\sum_{i=4}^{\infty} <L>_{F,i}g(i),
\ee
where $g(i)$ is the probability distribution of {\it i-cells} in the base. As
an approximation one can assume that the base network is hexagonal
 and obtain:
 \be\l{32}
<L>_{6,F}=<L>_F=\frac{F-7}{7}.
\ee
So the average volume (Lewis law in three dimensions) of
{\it F-cells} will than be:
\be\l{33}
<v>_F=S\frac{F-7}{7},
\ee
where $S$ is the average area of cells in the base.
Also, as second example, we assume that the network of base
is Poisson network with uniform density, which has explained in
previous section(eq.(\r{11})), so that eq.(\r{31}) will become:
\be\l{34}
<L>_F=\frac{1}{3}\sum_{i=4}^{\infty}<L>_{F,i}(\frac{2}{3})^{i-4}
=0.155F-1.
\ee
Therefore by making use of eq.(\r{13}) we can obtain the overage volume of
$\it {F-cells}$ in this model:
\be\l{35}
<v>_F=\frac{d_0}{3}(i-3)(0.155F-1).
\ee
From the formula for $ g_i(F)$  various moments can be calculated:
\bea\l{36}
<F>_i&=&2i+2,\\
<F^2>_i&=&5i^2+9i+4,
\eea
and
\bea\l{37}
<F>&=&2<i>+2=14,\\
<F^2>&=&5<i^2>+9<i>+4=268.
\eea

{\section {Poisson Networks With Non-Uniform Density}}

A little modification of the previous formulas is necessary
 to study the properties of a network where the density of points (the average
 distance between points to be denoted by $\lambda$, in the following)
 in different columns are different. Instead of eqs.(\r{3}) and (\r{4}) we  have:
 \be\l{40}
 P_\alpha(n)=(\frac{L}{\lambda_\alpha})^{n}\frac{1}{n!}
 e^{-\frac{L}{\lambda_\alpha}},
 \ee
\bea\l{41}
\Phi_{0}(L, n_1, n_2, \ldots, n_k)&=&P_{1}(n_1)P_{2}(n_2)\ldots P_{k}(n_k),
\nonumber \\
&=&\prod_{\alpha=1}^{k} (\frac{L}{\lambda_\alpha})^{n_\alpha}
\frac{1}{n_\alpha!}e^{-\frac{L}{\lambda_\alpha}},
\eea
respectively. By defining $\Psi(L,I)dL$ as before, we obtain:
\bea\l{42}
\Psi_0(L,I)&=&\sum'_{n_1+n_2+\ldots+n_k=I} \Phi(L, n_1, n_2, \ldots, n_k)
\frac{1}{\lambda_0}e^{-\frac{L}{\lambda_0}}dL,\nonumber \\
&=&\frac{L^{I}(\xi_0-\frac{1}{\lambda_0})^{I}}{\lambda_{0}{I!}}e^{-L{\xi_0}},
\eea
where $\xi_0=\frac{1}{\lambda_0}+\frac{1}{\lambda_1}+\frac{1}{\lambda_2}+
\ldots+\frac{1}{\lambda_k}$.
Now we study another probability density. We consider a cell of length
$L $ in which randomly located in the $C_n$th column (L is Poisson segment in
$C_n$th column). The cover length, $L_{c_{n+1}}$, of L is defined as the
smallest  length in $C_{n+1}$th column such as it is contain length L(i.e.
$L_{c_{n+1}} \geq L$)(fig(1c)). Here, we wish to obtain the probability
density of the cover length, $L_{c_{n+1}}$, whereas the main distribution
in the  columns of network is inhomogenous Poisson. So this new
probability destiny  must satisfy the following conditions:\\
1) In the constant length $L_{n+1}$ of L, there are m points of $C_{n+1}$th
non-uniform Poisson distribution, i.e.
\be\l{43}
P_{n+1}(m,L_{n+1})= (\frac{L_{n+1}}{\lambda_{n+1}})^m \frac{1}{m!}
e^{-\frac{L_{n+1}}{\lambda_{n+1}}}.
\ee
2) There are segments $\it l$ and $\it l'$ in $C_{n+1}$th non-uniform density
\bea\l{44}
P_{(n+1)}(\it l)&=& \frac{1}{\lambda_{n+1}}e^{-\frac{\it l}{\lambda_{n+1}}},\\
P_{(n+1)}(\it l')&=& \frac{1}{\lambda_{n+1}}e^{-\frac{\it l'}{\lambda_{n+1}}},
\eea
and
\bea\l{46}
L_{c_{n+1}}& =& L_{n+1}+\it{l+l'} ,\\
L&=&L_{n+1}+\it{t + t'},\\
\it{t}&\leq& \it{l} \leq L_{c_{n+1}}-L+\it{t}.
\eea
3)
\be\l{49}
\int_{L}^{\infty} {\cal P}_{n+1}(L_{c_{n+1}}, L)dL_{c_{n+1}} = 1.
\ee
where  $ {\cal P}_{n+1}(L_{c_{n+1}}, L)$ is the non-uniform probability
 density of cover length $L_{c_{n+1}}$.
 Therefore, the probability density of cover length in $C_{n+1}$th non-uniform
 Poisson distribution with above conditions is:\\
 ${\cal P}_{n+1}(L_{c_{n+1}},L,m) =$
 \be\l{50}
 Q\int_{\it t}^{L_{c_{n+1}}-L-{\it t}}
 \int P_{n+1}(\it l)P_{n+1}(\it l')P_{n+1}(m,L_{n+1})
 \delta{(L_{c_{n+1}}-L_{n+1}-\it{l-l'})}
d{\it l}d{\it l'},
\ee
where Q is a normalization factor. One can find probability density of cover
length for any m:
\be\l{51}
 {\cal P}_{n+1}(L_{c_{n+1}},L)=\sum_{m=0}^{\infty}
 {\cal P}_{n+1}(L_{c_{n+1}},L,m)=\frac{1}{{\lambda_{n+1}}^2}(L_{c_{n+1}}-L)
 e^{-\frac{1}{{\lambda_{n+1}}}(L_{c_{n+1}}-L)}.
 \ee
 We see that, this distribution is independent of $L_{n+1}$ and $\it t$.
 The average value of $(L_{c_{n+1}}-L)$ for fixed L in column $C_{n+1}$ is:
 \be\l{52}
 <(L_{c_{n+1}}-L)> = 2\lambda_{n+1}.
 \ee
 Note that, it is independent of L.
 In the case which we study the 3D poisson network, for a given segment $L$,
 there are two cover lengths, $L_{c_k}$ and $L_{c_j}$ in two independent poisson
 distribution, in which are neighbour with each other(fig1.d). It is possible
 that these two cover lengths  partly overlap. The probability density of the
 non-overlap length $|L_{c_k}-L_{c_j}|$ is  ${1\over\lambda_k}\exp({L_{c_k}
 -L_{c_j} \over \lambda_k})$, for $C_k$th column. The average of the
  non-overlap length in $C_k$th column is:
  \be\l{53}
  <|L_{c_k}-L_{c_j}|>=\lambda_k.
  \ee

\subsection{Statistical Properties of Non-Uniform Density}
{\Large{3.1.1 Two Dimensional Networks}}

In a two dimensional network every column say $C_{n}$ has two neighbours,
$C_{n-1}$ and $C_{n+1}$. The number of sides of {\it i-cell}
 in column $C_n$, is
$i=4+I$, where $I=n_{n-1}+n_{n+1}$ and $n_{n-1}+n_{n+1}$ are the number of
vertices contributed by the cells in adjacent columns. Then for the distribution
of {\it i-cells} of length between $(L $ and $ L+dL)$ in column $C_{n}$, we obtain:
\be\l{54}
g_n(L,i)= \frac{L^{i-4}(\xi_n-\frac{1}{\lambda_n})^{i-4}}{\lambda_n{(i-4)}!}
e^{-L\xi_n},
\ee
where $\xi_n=\frac{1}{\lambda_{n-1}}+\frac{1}{\lambda_{n}}+\frac{1}
{\lambda_{n+1}}$. Hence, the distribution of {\it i-cells}
with any arbitrary length in column $C_n$ is:
\be\l{55}
g_n(i)=\int_{0}^{\infty} g_n(L,i)dL= \frac{1}{\lambda_n\xi_n}
(1-\frac{1}{\lambda_n\xi_n})^{i-4},
\ee
clearly this distribution function is normalized.
From eq.(48) and (49) we also obtain the average height of $\it{i-cells}$ in
column $C_n$:
\be\l{56}
<L>_{n,i}= \int_0^{\infty}\frac{L g_n(L,i)dL}{g_n(i)}=
\frac{i-3}{\xi_n}.
\ee
So that for the Poisson networks with non-uniform density in one direction
($\it y$ direction)(fig1) Lewis law (average area of $\it{i-cells}$)
in column $C_n$ is:
\be\l{57}
<A>_{n,i} = \frac{d_0(i-3)}{\xi_n}.
\ee
One can now evaluate various moments of the distribution $g_{n}(i)$. In all the
 following calculations, we use the following formulas for geometric sum:
 \bea\l{58}
 \sum_{k=0}^{\infty} x^{k}& =& \frac{1}{1-x}, \\
 \sum_{k=0}^{\infty} kx^{k}& =& \frac{x}{(1-x)^{2}},\\
\sum_{k=0}^{\infty} k^{2}x^{k}& =& \frac{x (1+x)}{(1-x)^{3}}.
\eea
A simple calculation shows that:
\bea\l{61}
<i>_{n}&=&\sum_{i=4}^{\infty} ig_{n}(i)=3+\lambda_{n}\xi_{n}=4+\lambda_{n}
(\frac{1}{\lambda_{n-1}}+\frac{1}{\lambda_{n+1}}),\\
<i^2>_n&=&\sum_{i=4}^{\infty} i^2g_{n}(i)
=2(\lambda_{n}\xi_{n})^2+5\lambda_{n}\xi_{n}+9,
\eea
 where $<i>_{n}$ is the average number of {\it i-cells} in column $C_n$.
The variance turns out to be equal to:
\be \l{63}
<i^2>_n-<i>_n^2=(\lambda_{n}\xi_{n})(\lambda_{n}\xi_{n}-1).
\ee
In order to find the moments $<i>$ and $<i^2>$ in the whole lattice and not in a
particular column $C_n$, we replace the average over
the cells in a network, by an average over the densities in an ensemble of
networks. Thus from eqs.(\r{57}), (\r{61}) and (\r{62}) we have :
\bea\l{64}
\ll A \gg_i&=&d_0(i-3)<\frac{1}{\xi_n}>,\\
\ll i \gg&=&4+<\lambda_{n}(\frac{1}{\lambda_{n-1}}+\frac{1}
{\lambda_{n+1}})>,\\
\ll i^2 \gg&=&2<(\lambda_{n}\xi_{n})^2>+5<\lambda_{n}\xi_{n}>+9.
\eea
Where the averages on the right hand side are performed
with a suitable distribution $$P(\lambda_{1}, \lambda_{2}, \ldots , \lambda_{N}).$$
For simplicity in the following we assume that this distribution, is a symmetric
distribution, i.e:
\be\l{67}
P(\lambda_{1}, \lambda_{2}, \ldots , \lambda_{N})=
 P(\lambda_{1'}, \lambda_{2'}, \ldots , \lambda_{N'}),
 \ee
where the primed indices are a permutation of the unprimed ones.
 Clearly this minor restriction allows our conclusions to be valid for a very
 large class of probability distributions.
Finally, we consider Aboav-Weaire law for non-uniform Poisson distribution.
By using of eqs.(\r{22}),(\r{23}) and (\r{52}), we obtain:
\be\l{68}
<V_1 +V'_1>_{n+1} = \frac{<L_{c_{n+1}}-L_a>_{n+1}}{\lambda_{n+1}}=2,
\ee
\be\l{69}
<V''_1>_{n+1}=  \frac{<L_{c_{n+1}}>_{n+1}}{\lambda_{n+1}}+ \frac{<L_a>_{n+1}}
{\lambda_{n+1}} = 2 + \frac{1}{\lambda_{n+1}}<L_a>_{n+1}.
\ee
Combining eqs.(\r{24}), (\r{68}) and (\r{69}), we have:
\be\l{70}
<im_i> = 16+4i + \frac{<L_a>_{(n+1),i}}{\lambda_{n+1}} +
\frac{1}{\lambda_{n-1}}<L_a>_{(n-1),i}.
\ee
By using of eqs.(\r{57}) and (\r{70}), obtain:
\be\l{71}
<im_i> = 16 + 4i +(i-3)(\frac{1}{\lambda_{n+1} \xi_{n+1}} +
\frac{1}{\lambda_{n-1} \xi_{n-1}}).
\ee
At last we have:
\be\l{72}
\ll im_i \gg = 16 + 4i + (i-3)<\frac{1}{\lambda_{n+1} \xi_{n+1}} +
\frac{1}{\lambda_{n-1} \xi_{n-1}}> .
\ee
Where the last averaging are performed by a given distribution in
eq.(\r{67}). If we use of  uniform distribution we have
$\lambda_{n+1}\xi_{n+1} = \lambda_{n-1}\xi_{n-1} = 3$, so that:
\be\l{73}
\ll im_i \gg = 14 + \frac{14}{3}i,
\ee
we see that this result is the same in eq.(\r{26})

{\Large 3.1.2 Three-Dimensional Networks}

Consider a 3-dimensional laminated Poisson network, based on an arbitrary
2-dimensional network, and a particular ${\it i-cell}$ is the base. I label
the column with this base by $C_{i,0}$. The number of faces in a
cell in the column above this {\it i-cell} is
\be\l{74}
F = i + 2 +J,
\ee
where $J=n_1+n_2+ \ldots + n_i$, and $n_k$ is the number of additional lateral
 faces which results from the cells in the adjacent $k-$column (fig.3).
Let $f_i(J)$ be the fraction of the $\it{F-cells}$ in column $C_{i,0}$.
In formula (\r{41}), one should now replace $k$ by i, and I by J to obtain:
\be\l{75}
\Psi_0(L,F,i)=\frac{L^{F-i-2}(\xi_0-\frac{1}{\lambda_0})^{F-i-2}}{\lambda_0
(F-i-2)!}
e^{-L\xi_0},
\ee
where $\xi_0=\frac{1}{\lambda_0}+\frac{1}{\lambda_1}+\frac{1}{\lambda_2}+
\ldots+\frac{1}{\lambda_i}$. Here $\lambda_0$ is the density of P-point distribution in the column
$C_{i,0}$ and $\lambda_\alpha$ ($\alpha=1,2,\ldots,i$) are the density of
 P-point distribution in the adjacent columns. $\Psi_0(L,F,i)dL$ is the
probability of finding an {\it F-cell} in column $C_{i,0}$, whose height is
between $L$ and $L+dL$. Integrating over $L$, one obtains
\be\l{76}
\Psi_0(F,i)=\frac{1}{\lambda_0\xi_0}(1-\frac{1}{\lambda_0\xi_0})^{F-i-2},
\ee
this is the probability of finding an {\it F-cell} in column $C_{i,0}$.
The average height of $\it{F-cells}$ with $\it{i-cell}$ base in column $C_0$
is:
\be\l{77}
<L>_{0,i,F}= \int_0^{\infty} \frac{L\Psi_0(F,L,i)dL}{\Psi_0(F,i)}=
\frac{F-i-1}{\xi_0},
\ee
hence, the average height of $\it{F-cells}$ with any base in column $C_0$
\be\l{78}
<L>_{0,F}=\sum_{i=4}^{\infty} <L>_{0,i,F}g_0(i).
 \ee
 Here, we can calculate eq.(\r{78}) for three different base networks. Assume
 that the network  of base is:\\
 1) hexagonal
 \be\l{79}
 <L>_{0,F}= \frac{F-7}{\xi_0},
 \ee
 2) uniform Poisson network
 \be\l{80}
<L>_{0,F}=\frac{1}{3}\sum_{i=4}^{\infty}(\frac{F-i-1}{\xi_0})
(\frac{2}{3})^{i-4},
\ee
3) non-uniform Poisson network
\be\l{81}
<L>_{0,F}=\sum_{i=4}^{\infty}(\frac{F-i-1}{\xi_0})g_0(i),
\ee
note that the non-uniform Poisson distribution in horizontal columns (2D Poisson
network) and vertical columns are independent. Therefore  we can easily
calculate the final averaging in eq.(\r{74}).
One can now find various moments of this distribution:
\bea\l{82}
<F>_{i,0}&=&\lambda_0\xi_0+i+1,\\
<F^2>_{i,0}&=& (i+1)^2+ (2i+1)(\lambda_0\xi_0)+2(\lambda_0\xi_0)^2.
\eea
By the same averaging procedure as in the two dimensional case one can obtain:
 \bea\l{84}
 \ll L \gg_{F,i}& =& (F-i-1)<\frac{1}{\xi_0}>, \\
 \ll F \gg_i&=&<\lambda_0\xi_0>+i+1,\l{85}\\
\ll F^2 \gg_i&=& (i+1)^2+ (2i+1)<\lambda_0\xi_0>+2<(\lambda_0\xi_0)^2> \l{86}.
\eea
The later averages are performed over the density distributions eq.(\r{67}).
Note that for the distribution with uniform density in vertical
columns one has: $\lambda_0\xi_0=i+1$  and the eqs.(\r{84}), (\r{85}) and
 ({86}) reduce to:
\bea \l{87}
\ll L \gg_{i,F}&=& \frac{F-i-1}{i+1},\l{88}\\
\ll F \gg_i&=&2i+2,\l{89}\\
\ll F^2 \gg_i&=&(i+1)(5i+4),\l{90}
\eea
which are in agreement with equations (\r{30}), (\r{36}), (\r{37}).\\
In this section, we derive the 3D Aboav - Weaire law in non-uniform
Poisson network. Consider a 3D cell $a$ with $F$ faces, height $L_a$
and a base with $i$ edges in reference column, $C_0$.
 We want find the total number of faces in cells adjacent to $a$
with fixed $F$ and $i$. This quantity is $(Fm_F)_{0,i}$.
The number lateral walls in column $k$ is $m_i$. So that
the total number of faces, which obtain from P-points within $L_a$ in
first - neighbour columns to $a$ is:
\be\l{90}
J = \sum_{k=1}^i J_k = \sum_{k=1}^i{\large [}(m_i+2)(n_k+1) + (V_k+1)
+(V'_k+1){\large ]},
\ee
and, the total number of faces, which obtain from P-point within second-
neighbour columns to $a$ is:
\be\l{91}
J' = \sum_{k=1}^{im_i-3i} V'_k = (im_i -3i)<V'_k>,
\ee
where, $(im_i -3i)$ is the number of vertical walls of the adjacent cells that
contact neighbours, and also there are another faces which obtain from
P-points in pairs of adjacent column $k$, $j$ that are adjacent to each other.
So the total number of extra faces of such adjacenceis is
\be\l{92}
J'' = 2\sum_{k=1}^i [(n_k+1) +{<|L_{c_k} -
 L{c_j}|>\over \lambda_k} ],
\ee
where the number 2 is due to two distributions are overlap in each
vertical wall.
Clearly one can obtain $(Fm_F)_{0,i}$ as
\be\l{93}
<Fm_F>_{0,i} = 28 + J+ J'+J''.
\ee
Hence the number 28 comes from the average of faces of two cells adjacent to
$a$ in the  same reference column, $C_0$. By substituting eqs. (\r{90}),
 (\r{91}), (\r{92}) in (\r{93}), and using of(\r{52}), (\r{53}), (\r{68}),
   (\r{69}) and (\r{77}), we have
\be\l{94}
<Fm_F>_{0,i} = m_i[(1+\frac{i}{\xi_0})F + (2-\frac{(i+1)}{\xi_0})i -2]
+[4-\frac{3i}{\xi_0}]F + \frac{3i(i+1)}{\xi_0} +20,
\ee
and at last, we have
\bea\l{95}
\ll Fm_F \gg_i& =& m_i\{[i+i<{1\over \xi_0}>]F+[2-(i+1)
<{1\over \xi_0}>]i-2\}\nonumber\\
&+&\{4-3i<{1\over \xi_0}>\}F + 3i(i+1)<{1\over \xi_0}> +20.
\eea
Hence the double average are performed by a symmetric distribution
such as (\r{70}). It is remarkable that these averaging are performed
with respect to those distributions in vertical columns(for example z
direction), and are independent of distributions in base network.
So that if we use of non-uniform poisson network as base network, we have
two averaging, which are independent.
{\section {Conclusion}}

My main conclusion is that Lewis and Aboav- Weaire laws apply to Poisson
networks. I have shown that by more randomizing these networks; (i.e. in
the sense of random density distribution in different columns) one can
still obtain the above mentioned laws. In fact the most features of these
networks have in common with real cellular structures. But the only
property of real systems in which all the angles around any vertex are equal
is obviously absent in these networks. So we see that these model satisfy
Lewis and Aboav-Weaire laws with the coefficients which are slightly different
from the results obtained in [1,9]. Therefore despite this modification the
main features of this model is still unaffected and one can derive the
statistical and topological properties of random cellular structures from
this  modified model.\\

\end {document}